\documentclass[sigconf]{acmart}

\usepackage[inline]{enumitem}
\usepackage[binary-units]{siunitx}
\usepackage{tikz}
\usetikzlibrary{arrows,positioning,shapes.geometric}
\usepackage{smartdiagram}
\usepackage{listings}
\newcommand{\sepnum}[1]{\num[group-separator={,}]{#1}}
\usepackage{caption}
\usepackage{subcaption}
\usepackage{balance}

\AtBeginDocument{%
  }

\copyrightyear{2022}
\acmYear{2022}
\setcopyright{acmcopyright}
\acmConference[SSDBM 2022]{34th International Conference on Scientific and Statistical Database Management}{July 6--8, 2022}{Copenhagen, Denmark}
\acmBooktitle{34th International Conference on Scientific and Statistical Database Management (SSDBM 2022), July 6--8, 2022, Copenhagen, Denmark}
\acmPrice{15.00}
\acmDOI{10.1145/3538712.3538723}
\acmISBN{978-1-4503-9667-7/22/07}

\acmSubmissionID{52}

\begin{document}

\title{SciTS: A Benchmark for Time-Series Databases in Scientific Experiments and Industrial Internet of Things}

\author{Jalal Mostafa}
\orcid{0000-0003-2857-7816}
\affiliation{%
    \institution{Karlsruhe Institute of Technology}
    \city{Karlsruhe}
    \country{Germany}
}
\email{jalal.mostafa@kit.edu}

\author{Sara Wehbi}
\affiliation{%
    \institution{American University of Beirut}
    \city{Beirut}
    \country{Lebanon}
}
\email{saw06@mail.aub.edu}

\author{Suren Chilingaryan}
\orcid{0000-0002-2909-6363}
\affiliation{
    \institution{Karlsruhe Institute of Technology}
    \city{Karlsruhe}
    \country{Germany}
}
\email{suren.chilingaryan@kit.edu}

\author{Andreas Kopmann}
\orcid{0000-0002-2362-3943}
\affiliation{
    \institution{Karlsruhe Institute of Technology}
    \city{Karlsruhe}
    \country{Germany}
}
\email{andreas.kopmann@kit.edu}

\newcommand{\scits}{\textit{SciTS}}

\renewcommand{\shortauthors}{Mostafa et al.}

\begin{abstract}
    Time-series data has an increasingly growing usage in Industrial Internet of Things (IIoT) and large-scale scientific experiments. Managing time-series data needs a storage engine that can keep up with their constantly growing volumes while providing an acceptable query latency. While traditional ACID databases favor consistency over performance, many time-series databases with novel storage engines have been developed to provide better ingestion performance and lower query latency. To understand how the unique design of a time-series database affects its performance, we design \scits{}, a highly extensible and parameterizable benchmark for time-series data. The benchmark studies the data ingestion capabilities of time-series databases especially as they grow larger in size. It also studies the latencies of 5 practical queries from the scientific experiments use case. We use \scits{} to evaluate the performance of 4 databases of 4 distinct storage engines: ClickHouse, InfluxDB, TimescaleDB, and PostgreSQL.
\end{abstract}



\maketitle
\section{Introduction}

The relational model of old Relational Database Management Systems (RDBMS) in addition to their robust implementations of the \textit{ACID} principles made them popular as general-purpose data stores. RDBMSs favor consistency over availability and performance which complicates scaling the system horizontally with efficiency in big data scenarios \cite{traditionaldbms}. As a result, new DBMSs were developed to relax some consistency constraints and provide better scalability and performance. Many new technologies, therefore, were introduced including
\begin{enumerate*}[label=(\arabic*)]
    \item wide-column stores e.g. \textit{Google Bigtable} \cite{bigtable}, \textit{Apache Cassandra} \cite{cassandra}, \textit{Apache HBase};
    \item key-value stores \textit{Amazon DynamoDB} \cite{dynamodb}, \textit{LevelDB}, and \textit{RocksDB};
    \item document-based stores \textit{AsterixDB} \cite{asterix}, \textit{ArangoDB}, and \textit{MongoDB} \cite{mongodb_sensors,mongodb_postgres,mongodb_bigdata};
    \item column-oriented stores e.g. \textit{Apache Druid} and \textit{ClickHouse} \cite{clickhouse_bigdata};
    \item graph stores \cite{graphdbs} e.g. \textit{Neo4j}.
\end{enumerate*}
However, the evolution of time-series applications in big data environments like large-scale scientific experiments, Internet of Things (IoT), IT infrastructure monitoring, industrial control systems, and forecasting and financial trends allowed the emergence of many Time-Series Databases (TSDB) technologies.

The emergence of TSDBs was motivated by the special characteristics of time-series data in comparison to other types of big data. Time-series data is: \begin{enumerate*}[label=(\arabic*)]
    \item indexed by its corresponding timestamps;
    \item continuously expanding in size;
    \item usually aggregated, down-sampled, and queried in ranges;
    \item and has very write-intensive requirements.
\end{enumerate*}
Different TSDBs developed distinct technologies to tackle these characteristics by designing storage engines that are capable of the heavy-write requirements and by accomodating indexing algorithms that provide low query latency. However, each distinct TSDB architecture yields a distinct performance.


This paper proposes \scits{} a new benchmark to study the performance of distinct TSDBs in the scenarios of scientific experiments and industrial IoT. The benchmark simulates heavy INSERT workloads as well as 5 queries inspired by the use case of time-series data in a scientific experiment. This effort is motivated by our own experiences to set up evaluation testbeds for TSDBs to accommodate them in the KArlsruhe TRItium Neutrino Experiment ({KATRIN}) \cite{katrin_design}. \scits{} can simulate any workload by parameterizing concurrency, cardinality, and size of batches while considering best performance practices for each workload type. Unlike other benchmarks, it introduces a new workload "Scaling Workload" to understand the performance of TSDBs as the data inside the database grows larger. In addition, \scits{} also collects usage of the system resources like CPU and memory usage.

As mentioned above, our benchmark is the product of hours of research in testing and evaluating TSDBs for scientific experiments. Based on our experiences, we gathered a list of requirements for a good TSDB benchmark:
\begin{itemize}
    \item \textbf{Customizability \& Extensibility:} an easy and highly extensible interface to generate and test different types of {INSERT} workloads;
    \item \textbf{Practical Queries:} queries from practical and real-life environments e.g. range queries, out-of-range queries, and more complex queries like aggregation and down-sampling queries;
    \item \textbf{Scalability Performance:} the ability to understand the performance of a TSDB as its size grows larger;
    \item \textbf{System Monitoring:} the capability to monitor the usage of system resources.
\end{itemize}
Existing TSDB benchmarks only support a limited set of queries or do not reflect on the scalability performance of a TSDB \cite{iotdbbenchmark,tsbenchmark,ycbsts,tsbs}. Our benchmark builds on previous efforts by providing queries from real-life scenarios, specifically scientific experiments, and by giving insights into the scalability performance of TSDBs.

To evaluate our benchmark, we choose 3 TSDBs of three distinct storage engines: \textit{InfluxDB} \cite{influx} to represent TSDBs based on LSM trees, \textit{TimescaleDB} \cite{timescale} to represent TSDBs based on traditional RDBMSs, and \textit{ClickHouse} \cite{clickhouse_bigdata} to represent column-oriented OLAP-based TSDBs. We compare the performance of the three chosen TSDBs to \textit{PostgreSQL} as a reference line for traditional RDBMS. In summary, our contributions are:
\begin{itemize}
    \item A new extensible and parameterizable benchmark for TSDBs that focuses on heavy-write operations and query scenarios in addition to system resource monitoring.
    \item Insights on the performance of some state-of-the-art TSDBs using our benchmark and their underlying indexing and storage techniques.
    \item A performance analysis and comparison of distinct TSDB storage engines.
\end{itemize}

The rest of this paper is divided as follows: Section \ref{sec:scenario} explains the requirements of scientific experiments and industrial IoT that inspire our benchmark workloads. Section \ref{sec:workloads} specifies the data ingestion and query workloads that we use in this paper to understand the performance of TSDBs. The architecture and the components of \scits{} are discussed in Section \ref{sec:benchmark}. The experimental setup and the database servers configurations are discussed in Section \ref{sec:setup}. Using our benchmark, the performance of ClickHouse, TimescaleDB, PostgreSQL, and InfluxDB is reflected in Section \ref{sec:results}. Section \ref{sec:relatedwork} lists related work. We conclude in Section \ref{sec:conclusion}.

\section{Scientific Experiments Scenario}
\label{sec:scenario}
TSDBs have found a very tight acceptance in scientific experiments. Thousands of sensors in these experiments continuously read important timely metrics that contribute to the operation and to the scientific value of the experiment e.g. ion monitoring, electric current and voltage, magnetic field, temperature, etc. To store the values issued by these sensors, a data store that is efficiently designed for write operations is needed to cover the high data ingestion rate caused by a large number of installed sensors and the requirement to sustain
relatively high sampling rates. RDBMSs fail in such scenarios because they are not optimized for heavy-write operations and cannot scale out very well to provide high availability and to protect the cluster from having a single point of failure \cite{traditionaldbms}. In addition, traditional RDBMSs use query engines that have very limited support to offload some of the data visualization and analysis tasks to the database itself, e.g. grouping by time ranges is much simpler with TSDBs' builtin functions. On the contrary, specialized TSDBs offer horizontal scalability for writing operations and very good support for the analysis and the visualization applications of time-series data as well as very decent performance to ingest the data of a very large number of sensors that are typically sampled at frequencies ranging from \SI{10}{\hertz} to \SI{0.1}{\hertz} and even higher in special cases involving very fast processes particularly related to disaster protection, e.g. magnet quench detection systems \cite{katrin_design,lhc,panda_design,star}. Consequently, the characteristics of TSDBs make them interesting candidates to store sensor readings in scientific experiments.

Our paper takes scientific experiments such as {KATRIN} as a use case to evaluate TSDBs. The queries we propose in \scits{} are extracted from data management systems of many scientific experiments after hours of log analysis to understand what data and information physicists are trying to look at. Our benchmark parameters are inspired by the number of sensors and the sampling rates of these experiments and particularly {KATRIN}. Although we design our benchmark around scientific experiments, it is highly flexible and can represent any kind of workload particularly industrial IoT.

\section{Benchmark Workloads}
\label{sec:workloads}
Based on the scenario described in Section \ref{sec:scenario}, we propose eight types of benchmark workloads (3 data ingestion workloads and 5 query workloads) for time-series databases.

\subsection{Data Ingestion Workloads}
Ingestion workloads are designed for scientific experiments and industrial IoT but they are very flexible and can be extended to any time-series scenario. Extensions to \scits{} ingestion workloads is possible by changing three relevant parameters: concurrency i.e. number of clients, size of data batches, and cardinality i.e. number of sensors. Using these parameters, the user of the benchmark can create any workload scenario. For our study, we introduce 3 data ingestion workloads focusing on batching, concurrency, and scaling.

\paragraph{Batching Workload}
Understanding the performance of databases under different batch sizes helps in evaluating how they react to small and big batches. This evaluation is important to decide how to use the database e.g. what is the most suitable batch size for a specific database? or how far can we go in the size of inserted data batches? For this workload, we consider varying the batch size while using only one database client. We consider batch sizes as small as 1000 points per batch and as large as 100000 points per batch. We study the latency brought in by inserting data of different batch sizes.

\paragraph{Concurrency Workload}
Any practical use of databases in industrial IoT and scientific
instrumentation includes using numerous clients that are responsible for
reading sensor data from multiple devices and writing the results into
the database in batches. The concurrency workload tests the performance of TSDBs by varying the number of clients and monitoring the ingestion rate of the database as well as the usage of system resources.

\paragraph{Scaling Workload}
Different databases have different backends that use memory and storage resources in distinct ways. While some databases may support higher ingestion rates than others, it is important to study the performance of the database as data grows larger. The goal of this workload is to study the performance of TSDBs as they grow in size over time. It involves collecting and studying the usage of system resources to understand the impact of system resources on data ingestion.

\subsection{Queries Workload}
\scits{} proposes five queries that are inspired by the {KATRIN}'s data visualization and analysis workloads. The queries focus on returning raw, aggregated, or down-sampled data of one or more sensors. We define data aggregation as summarizing a duration of time using one single value e.g. average, standard deviation, etc. On the other hand, we define down-sampling as the practice of summarizing the sensor's data on a time interval basis using a statistical function to create a new time-series of summarized intervals.

Assuming the data is defined using the constructs of a relational table, the table schema would be \textit{(time\_field, sensor\_id, value\_field)}. We also assume that the function \textit{TRUNCATE} is a function that returns a list of
time intervals of a specified length e.g. \textit{TRUNCATE('1min', time\_field)} will return a list of time-intervals where each item represents a 1-minute of data using the column \textit{time\_field}. Using this schema, the queries and their SQL equivalents can be described as follows:
\begin{enumerate}[label=\textbf{(Q\arabic*)}]
    \item Raw Data Fetching: Get the raw values of one or more sensors over a duration of time. It is used to visualize and analyze data of specific sensors.

          \begin{lstlisting}
SELECT *
FROM sensors_table
WHERE time_field > ?
AND time_field < ?
AND sensor_id = ANY(?, ?, ?, ...)
\end{lstlisting}

    \item Out of Range Query: Get the intervals over a duration of time where the value of a specific sensor was out of a defined range. It is used to detect when the sensor was acting abnormally in a specific interval of time.

          \begin{lstlisting}
SELECT TRUNCATE(period, time_field)
AS interval, MAX(value_field),
MIN(value_field)
FROM sensors_table
WHERE time_field >= ?
AND time_field <= ?
AND sensor_id = ?
GROUP BY interval
HAVING MIN(value_field) < ?
OR MAX(value_field) > ?
\end{lstlisting}

    \item Data Aggregation: Represent the data of one or more sensors over a specific duration of time using one aggregated value of an aggregation function denoted by \textit{agg\_func} e.g. the standard deviation, the mean, etc.

          \begin{lstlisting}
SELECT agg_func(value_field)
FROM sensors_table
WHERE time_field >= ?
AND time_field <= ?
AND sensor_id = ANY(?, ?, ?, ...)
\end{lstlisting}

    \item Data Down-Sampling: down-sample one or more sensors using a specific sampling function denoted by \textit{agg\_func} over a duration of time.

          \begin{lstlisting}
SELECT TRUNCATE(period, time_field)
AS interval, sensor_id,
agg_func(value_field)
FROM sensors_table
WHERE time_field >= ?
AND time_field <= ?
AND sensor_id = ANY(?, ?, ?, ...)
GROUP BY interval, sensor_id
\end{lstlisting}

    \item Operations on Two Down-sampled Sensors: Down-sample the data of two sensors over a duration of time and using the function \textit{agg\_func}, then compare the results using the function \textit{comp\_func}. A use case of this query is comparing the data of two down-sampled sensors using value subtraction.

          \begin{lstlisting}
SELECT Sensor1.period,
comp_func(Sensor1.val, Sensor2.val)
FROM
  (SELECT TRUNCATE(period, time_field)
   AS interval,
   agg_func(value_field) AS val
   FROM sensors_table
   WHERE time_field >= ?
   AND time_field <= ?
   AND sensor_id = ANY(?, ?, ?, ...)
   GROUP BY interval)Sensor1
INNER JOIN 
  (SELECT TRUNCATE(period, time_field)
   AS interval,
   agg_func(value_field) AS val
   FROM sensors_table
   WHERE time_field >= ?
   AND time_field <= ?
   AND sensor_id = ANY(?, ?, ?, ...)
   GROUP BY interval)Sensor2
ON Sensor1.period = Sensor2.period
\end{lstlisting}
\end{enumerate}

\section{The Benchmark Architecture}
\label{sec:benchmark}
\begin{figure}[ht]
    \centering
    \Description{The SciTS Process Flow consists of 6 components: the configuration, the workload generator, the database abstraction layer, parallel clients, system resource monitoring, and the performance metrics logger.}
    \includegraphics[width=0.46\textwidth]{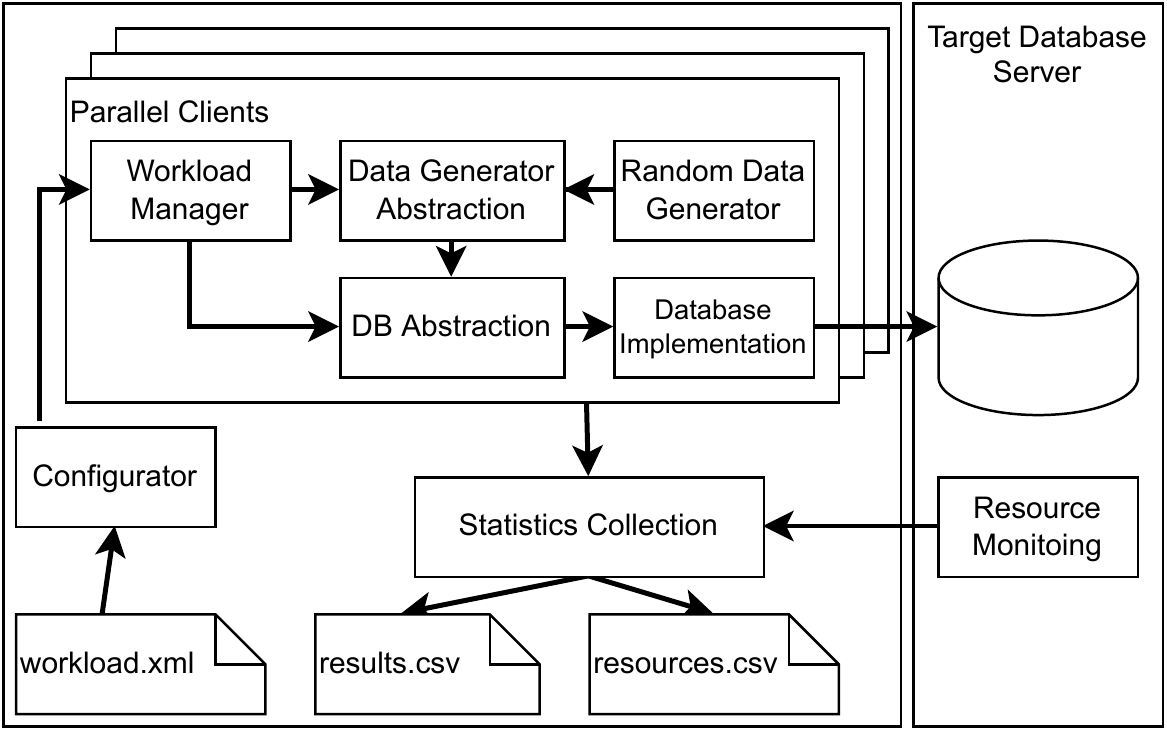}
    \caption{The Architecture and Process Flow of SciTS}
    \label{fig:scitsarch}
\end{figure}

This section provides an overview of the architecture of \scits{} and its design that supports the requirements discussed in Section \ref{sec:scenario}. \scits{} is an extensible configurable client-side benchmark that can work for any single node DBMS. \autoref{fig:scitsarch} shows the architecture and the control flow of \scits{}. The benchmark flow starts the configurator that reads the user's configurations and parameters from the workload definition file to create and launch a parallelized benchmark scenario. The configurator then creates the requested parallel clients. Each client operates a workload manager to create and submit workloads to the target database server. For ingestion workloads, the workload manager submits a request to the data generator abstraction layer to create sensor data. The generated sensor data is then passed to the database abstraction layer, an abstract interface that wraps the implementations of database clients. On the other hand, the parameters of query workloads are submitted directly to the database abstraction layer for execution. While executing the workloads, \scits{} asynchronously monitors the usage of the system resources on the target database server. The collected workload performance metrics and the system resources metrics are then recorded and persisted in separate files.

\scits{} is extensible through its abstract interfaces and resilient configurations. It abstracts database access, workloads, and data generations that are easy to extend for additional benchmark scenarios. For instance, \scits{} uses a random data generator by default, but additional data generators can be added by providing other implementations of the data generation abstraction interface. Similarly, additional queries and new database servers can be added by extending the relevant interfaces. Data ingestion workloads are extensible via benchmark specifications described in the workload definition file e.g. a concurrency workload in \scits{} is a data ingestion workload that vary the number of clients in its definition file and fix the batch size.

\subsection{Workload Definitions}

A \scits{} workload is a set of parameters in its XML configuration file in addition to information about the target database server and its connection specifications. Date and time span can be described in a workload definition to describe how sensors' timestamps are distributed over a specific period.

\autoref{tab:parameters} shows the user-defined parameters of \scits{}. In addition to generic parameters like \textit{TargetDatabase}, \textit{DaySpan}, and \textit{StartTime}, \scits{} defines parameters for each workload type. An ingestion workload is defined by parameterizing \scits{} using:
\begin{enumerate*}[label=(\arabic*)]
    \item \textit{ClientNumberOptions} to represent concurrency i.e. the number of database clients to insert records into the database,
    \item \textit{BatchSizeOptions} to configure the batch size to insert in one operation,
    \item and \textit{SensorNumber} to parameterize the cardinality of the database table by configuring a specific number of sensors.
\end{enumerate*}
For instance, a concurrency workload is defined by setting the \textit{ClientNumberOptions} to a set of number of clients to test with e.g. setting it to 1,2,4 means run the same workload with one database client, then two clients, then four clients in one atomic run without changing the configuration. The batching workload is another example where the user can similarly set \textit{BatchSizeOptions} to a set of batch sizes to test the database server with in one atomic run.

On the other hand, the user can specify in the configuration file what query he needs to execute using the \textit{QueryType} option. The five queries can be parameterized by choosing the queried time intervals (\textit{DurationMinutes} in \autoref{tab:parameters}) for, and by filtering on one or more sensors using the \textit{SensorsFilter} parameter. Down-sampling and aggregation queries are additionally parameterized by specifying aggregation or sampling interval. The benchmark uses the \textit{average} function to calculate aggregations. Other queries like out-of-range queries that require filtering on the \textit{value} column can be parameterized in the configuration file using the \textit{MinValue} and \textit{MaxValue} fields. To assess the results correctness, the user can repeat the same query with the same parameters as much as needed using the \textit{TestRetries} parameter.

\begin{table*}[ht]
    \caption{User-defined Parameters of \scits{} Workloads}
    \label{tab:parameters}
    \begin{tabular}{lp{9cm}c}
        \toprule
        Name                    & Description                                                           & Workload Type    \\
        \midrule
        TargetDatabase          & The type of the target database server e.g. InfluxDB, ClickHouse, etc & Ingestion/Query  \\
        DaySpan                 & Length of the whole time-series in the database table in days         & Ingestion/Query  \\
        StartTime               & Earliest timestamp to be stored into or retrieved from the database   & Ingestion/Query  \\
        BatchSizeOptions        & Size of batch to insert into table                                    & Ingestion        \\
        ClientNumberOptions     & Number of concurrent clients                                          & Ingestion        \\
        SensorNumber            & Number of sensors to simulate to represent cardinality                & Ingestion        \\
        QueryType               & An enum representing the query type e.g. Q1-Q5                        & Query            \\
        TestRetries             & How many times to repeat the query test                               & Query            \\
        DurationMinutes         & Length of time-series data in minutes                                 & Query (Q1 to Q5) \\
        AggregationIntervalHour & Length of time window to apply the down-sampling function on          & Query (Q3 to Q5) \\
        SensorsFilter           & A list of sensor IDs to filter on in the query                        & Query (Q1 to Q5) \\
        MaxValue                & The upper boundary of the sensor's value used in Q2                   & Query            \\
        MinValue                & The lower boundary of the sensor's value used in Q2                   & Query            \\
        \bottomrule
    \end{tabular}
\end{table*}

\subsection{Performance Metrics}
We evaluate the performance of data ingestion workloads by monitoring the latency taken to insert batches to the target database. We also consider the ingestion rate of the database (the sum of all inserted data points divided by the time it has taken to finish the insertion transaction). In scaling workloads, we consider a rolling ingestion rate where we resample the data on an one-minute interval basis then we calculate the ingestion rate for each of these intervals.

To evaluate query workloads, we consider the latency taken to execute and return the query results. We use the \textit{TestRetries} parameter to repeat the queries 1000 times then we study the samples' minimum, maximum, average, standard deviation, and 95\% percentile.

The benchmark monitors the usage of system resources of the server by using Glances \cite{glances}. In general, \scits{} monitors CPU (I/O wait, system, user, context switches), physical memory (used and cached memory), swap usage, disk I/O (read/write in bytes per second, count of I/O operations), and network usage (sent and received).

\subsection{The Implementation}
\scits{} is implemented using portable cross-platform C\#. \scits{}'s implementation is highly extensible using its abstraction layers and resilient configuration. The benchmark can support any database management system as long as the DBMS client is implemented using the database abstraction layer. The current implementations include \textit{ClickHouse} \cite{clickhouse_bigdata}, \textit{InfluxDB} \cite{influx}, \textit{PostgreSQL} \cite{pg}, and \textit{TimescaleDB} \cite{timescale}. We try to adopt best practices for each implementation to achieve the best possible performance, for instance: \textit{PostgreSQL} and \textit{TimescaleDB} bulk inserts are powered by PostgreSQL SQL \textit{COPY} statement that is optimized for high-performance ingestion rates with less locking and fewer indexing updates.

\scits{} implements a random data generator for data ingestion. The data generator generates timestamps incrementally based on the date and periods defined in the workload definition file. The granularity of the timestamps is configured also configurable in the workload definition file. Sensors' values are considered to be random values that are uniformly ranging between zero and the max value of a signed 32 bits integer.

\section{Experiments Setup}
\label{sec:setup}
We use two machines for our benchmarks: \textit{M1} \& \textit{M2} which work as a server and a client to perform our tests. Machine \textit{M1} is an enterprise-grade storage server that we use to host the database servers. It is equipped with an Intel Xeon CPU E5-1620 v2 @ \SI{3.70}{\giga\hertz} of 8 logical cores, \SI{32}{\giga\byte} DDR3 RAM, and 24 physical drives formatted with XFS and configured with RAID60 and connected using a \SI{6}{\giga\bit/\sec} SAS-2 controller. Machine \textit{M2} acts as the client machine. It is equipped with Intel Xeon CPU E5-2680 v3 @ \SI{2.50}{\giga\hertz} over 2 sockets of 48 logical cores in total, and \SI{512}{\giga\byte} DDR4 RAM in total. Both machines are connected over a \SI{1}{\giga\bit/\sec} Ethernet switch. We monitor the network traffic of both servers to make sure the connection is not saturated.

For our tests, we consider the following table schema for all databases where we store all measurements and data points in one table: \textit{(timestamp, sensor\_id, value)}. A record in this schema is represented by an 8 bytes timestamp of when the data point of the sensor was taken, an 8 bytes long integer as the ID of the corresponding sensor, and 8 bytes double-precision float. In all databases, we add indexes (a combined index) on the \textit{timestamp} field and
\textit{sensor\_id}, so it is faster to query for data points for a specific duration and specific sensors.

We use machine M1 for all of the database servers. Each server runs independently of the others while the others are all down. For all database servers, we use only one node. Evaluating the performance of a cluster of database server nodes is out of the scope of this paper.

We use the following database servers and configuration to allow the best possible performance:

\paragraph{ClickHouse} It is a column-oriented OLAP DBMS designed for high ingestion rates. ClickHouse's storage engine is called \textit{MergeTree} that writes the data directly to the table part by part to offer high-speed unrestricted data insertion. A background job then merges the parts. Data in ClickHouse can be stored sorted on the disk which allows using sparse indexing to locate data in the partitions quickly. We configure the database server to partition data every day. Each partition is then ordered by the table's primary key the tuple \textit{(timestamp, sensor\_id)}. Indices are defined on both of the fields: \textit{timestamp}, and \textit{sensor\_id}. We use ClickHouse v22.1.3.7 with its native TCP protocol and we set the following configurations: \textit{max\_server\_memory\_usage\_to\_ram\_ratio} to 0.9, \textit{index\_granularity} is 8192 rows, and \textit{async\_insert} is off.

\paragraph{InfluxDB} It is a TSDB that uses the \textit{Time-Structured Merge Tree} (TSM Tree), a storage engine that resembles Log-Structured Merge (LSM) trees \cite{lsmtree} in its design. Inserted data in TSM trees is written to a Write-Ahead Log (WAL) at first and copied to the cache while maintaining indexes in memory. The data is then persisted on the storage using immutable shards, each shard contains the data of a corresponding duration of time. An InfluxDB record consists of a timestamp, a value, and one or more tags. Tags are key-value pairs that are used to add data to the record. InfluxDB uses timestamps and tags for indexing. It uses per-type data compression algorithms e.g. ZigZag encoding for integers, the Gorilla algorithm \cite{gorilla} for float numbers, simple8b \cite{simple8b} for timestamp indexes, bit packing for booleans, and the snappy algorithm \cite{snappy} for strings. We use InfluxDB v2.1.1 and the \textit{Line} protocol to insert data and the server is set up with the following configuration: \textit{storage-wal-fsync-delay} is set to 0, \textit{storage-cache-max-memory-size} is set to \SI{1048}{\mega\byte}, and \textit{storage-cache-snapshot-memory-size} is set to \SI{100}{\mega\byte}.

\paragraph{PostgreSQL} It is an RDBMS that uses WAL to insert data. The WAL ensures the reliability of the data written to the database. It protects the data from power loss, operating system failure, and unanticipated hardware failures. We set up a PostgreSQL table with the previously discussed schema on one PostgreSQL v13.5 server and use B-Tree indexes on the fields \textit{timestamp} and \textit{sensor\_id} to find data quickly on a time range and for specific sensors. To optimize configurations for the host machine, the server is configured with \textit{pgtune} \cite{pgtune} with the following configurations: \textit{shared\_buffers} is \SI{7994}{\mega\byte}, \textit{maintenance\_work\_mem} is \SI{2047}{\mega\byte}, and \textit{max\_parallel\_workers} is 8 workers.

\paragraph{TimescaleDB} It is an extension of PostgreSQL. TimescaleDB benefits from the reliability and the robustness of PostgreSQL in addition to its SQL query engine. To solve the problem of always growing data, TimescaleDB uses hypertables that partition the data by the time column into several chunks. Each chunk is a standard PostgreSQL table. Standard SQL queries can be applied to the hypertable. This architecture handles time-series data better than traditional PostgreSQL. Indexing per chunk and chunks that can fit in the memory allows higher ingestion rates than traditional PostgreSQL. For low query latency, TimescaleDB uses age-based compression that transforms rows into a columnar format. Based on TimescaleDB recommendations, we set up a TimescaleDB v2.5.1 server with a \textit{hypertable} of a 12-hours chunking interval so chunks constitute no more than 25\% of the main memory. TimescaleDB compression is configured to compress row data into the columnar format every 7 days of data and to order the columnar data by \textit{timestamp} and \textit{sensor\_id}. The server is configured with the pgtune-based tool \textit{timescale-tune} with similar configuration as PostgreSQL

\section{Experiment Results}
\label{sec:results}
This section discusses the results and the analysis we did after applying \scits{} workloads to the target databases. For each of the experiments below, we consider the scientific experiments scenario with \sepnum{100000} sensors in total and in order to provide a realistic case of cardinality in the database.

\subsection{Data Ingestion}

\paragraph{Batching Workload} The goal of this workload is to understand how different database servers react to different batch sizes. We vary the batch size for each database then we measure the latency taken to insert each of these batches. For all databases and each of the batch sizes, we start from an empty database to keep the data of the experiments statistically independent as much as possible. We vary the batch size from \SI{1000}{records} until we reach \SI{100000}{records}, the maximum number of records {KATRIN} control system can have in a second.

\begin{figure}[ht]
    \centering
    \includegraphics[width=0.47\textwidth]{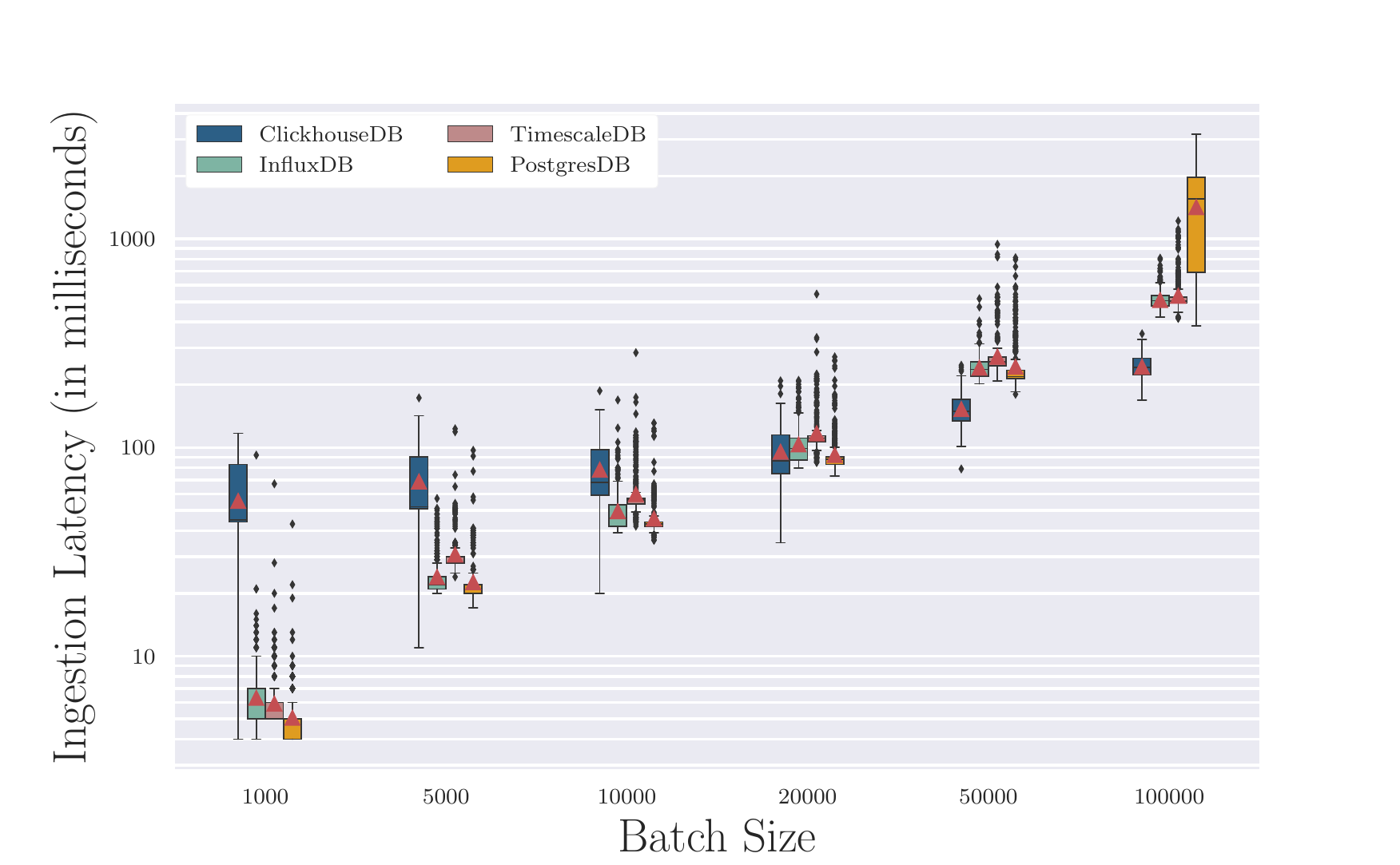}
    \caption{Batch Ingestion Latency as Function of Batch Size}
    \label{fig:batching}
    \Description{ClickHouse works better for larger batch sizes while other databases work better for smaller batch sizes. At batch size 20000, the databases show close performance. PostgreSQL show very high latency for batch size 100000.}
\end{figure}

\autoref{fig:batching} shows a box plot of the batch ingestion latencies and their mean values on a log scale as a function of the batch size for each of the target databases. Each box plot corresponds to the insertion of 500 batches into the target database. For batch sizes smaller than \sepnum{10000}, the traditional relational design of PostgreSQL performs better than time-series databases. Since ClickHouse's MergeTree writes directly to the storage, the latency produced by frequent write operations prevents ClickHouse from performing as well as other databases. For \sepnum{20000} data points in a batch, the four databases perform close to each other, and their means are in the range \SI{95}{\ms}--\SI{116}{\ms}. For huge batch sizes like \sepnum{50000} and \sepnum{100000}, ClickHouse outperforms all other databases.

TimescaleDB, InfluxDB, and PostgreSQL provide close performance for most of the batch sizes except in batch size \sepnum{100000} where PostgreSQL fails to handle very large data batches and the latency to insert one batch can reach more than \SI{3000}{\ms} while the chunks of TimescaleDB hypertables provide much better performance compared to traditional PostgreSQL.

\paragraph{Concurrency Workload} The goal of this workload is study the performance of the databases as the number of clients varies. For each of the target databases, we start from an empty table then we start varying the number of clients that are inserting data into the table. As we vary the number of clients, we calculate the total ingestion rate and check the CPU and the memory usages for each database. We choose a batch size of \sepnum{20000} since all targeted databases have a close ingestion latency as shown in \autoref{fig:batching}.

\begin{figure}[ht]
    \centering
    \includegraphics[width=0.47\textwidth]{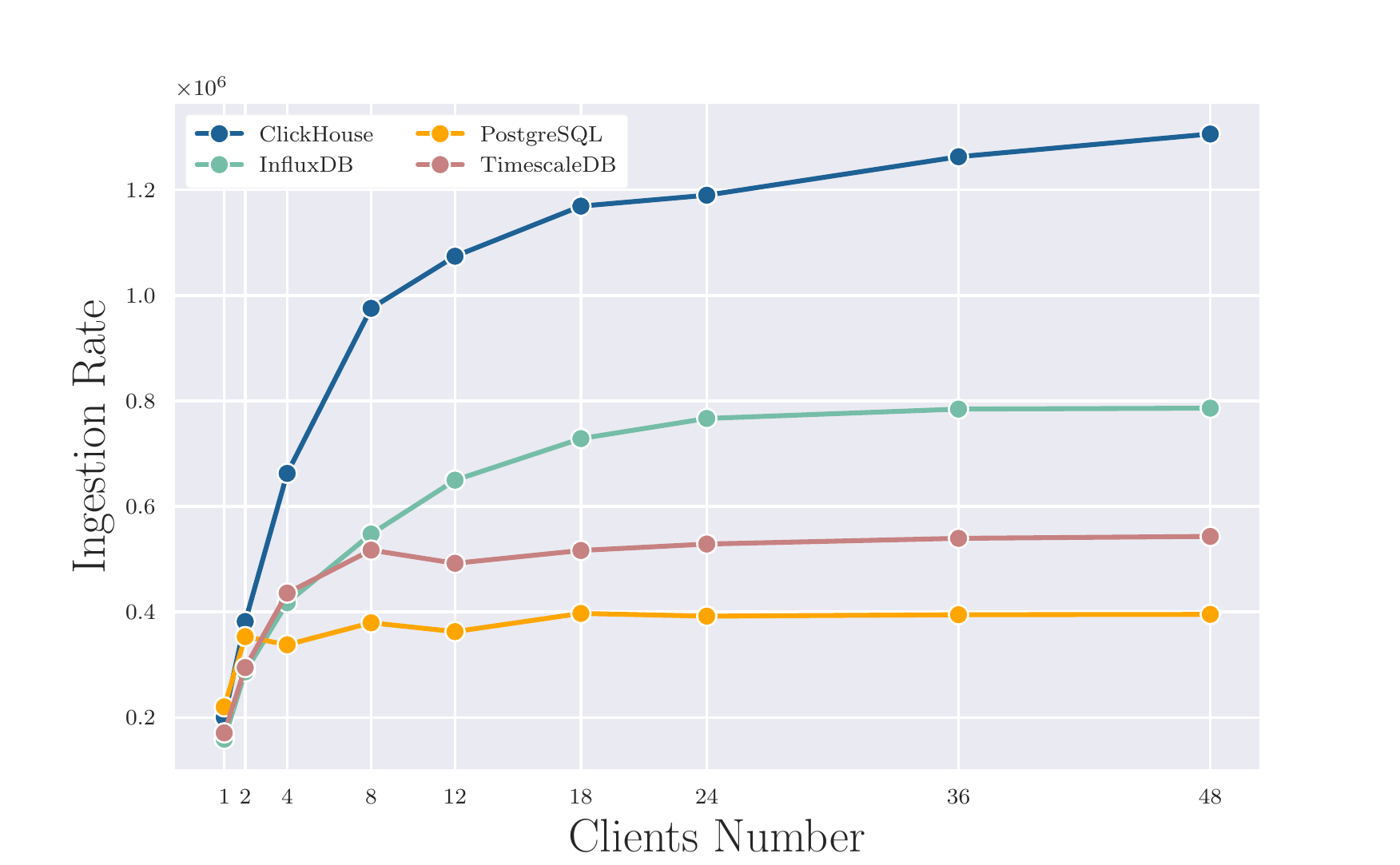}
    \caption{Ingestion Rate (in records per second) as Function of the Number of Concurrent Clients}
    \label{fig:clients}
    \Description{ClickHouse achieves the best ingestion performance where the ingestion rate can hit 1.3 million data points per second on average while using 48 clients. While ClickHouse shows an increasing performance with the increasing number of concurrent clients, other databases show some performance limits: InfluxDB is saturated with 24 clients and cannot achieve more than \sepnum{790000} points per second; TimescaleDB and PostgreSQL reach their peak performance at \sepnum{550000} and \sepnum{400000} respectively.}
\end{figure}

\autoref{fig:clients} shows the ingestion rate as a function of clients for each of the target databases. ClickHouse achieves the best ingestion performance where the ingestion rate can hit 1.3 million data points per second on average while using 48 clients. While ClickHouse shows an increasing performance with the increasing number of concurrent clients, other databases show some performance limits: InfluxDB is saturated with 24 clients and cannot achieve more than \sepnum{790000} points per second; TimescaleDB and PostgreSQL reach their peak performance at \sepnum{550000} and \sepnum{400000} respectively.

\begin{figure}[ht]
    \begin{subfigure}[b]{0.47\textwidth}
        \centering
        \includegraphics[width=\textwidth]{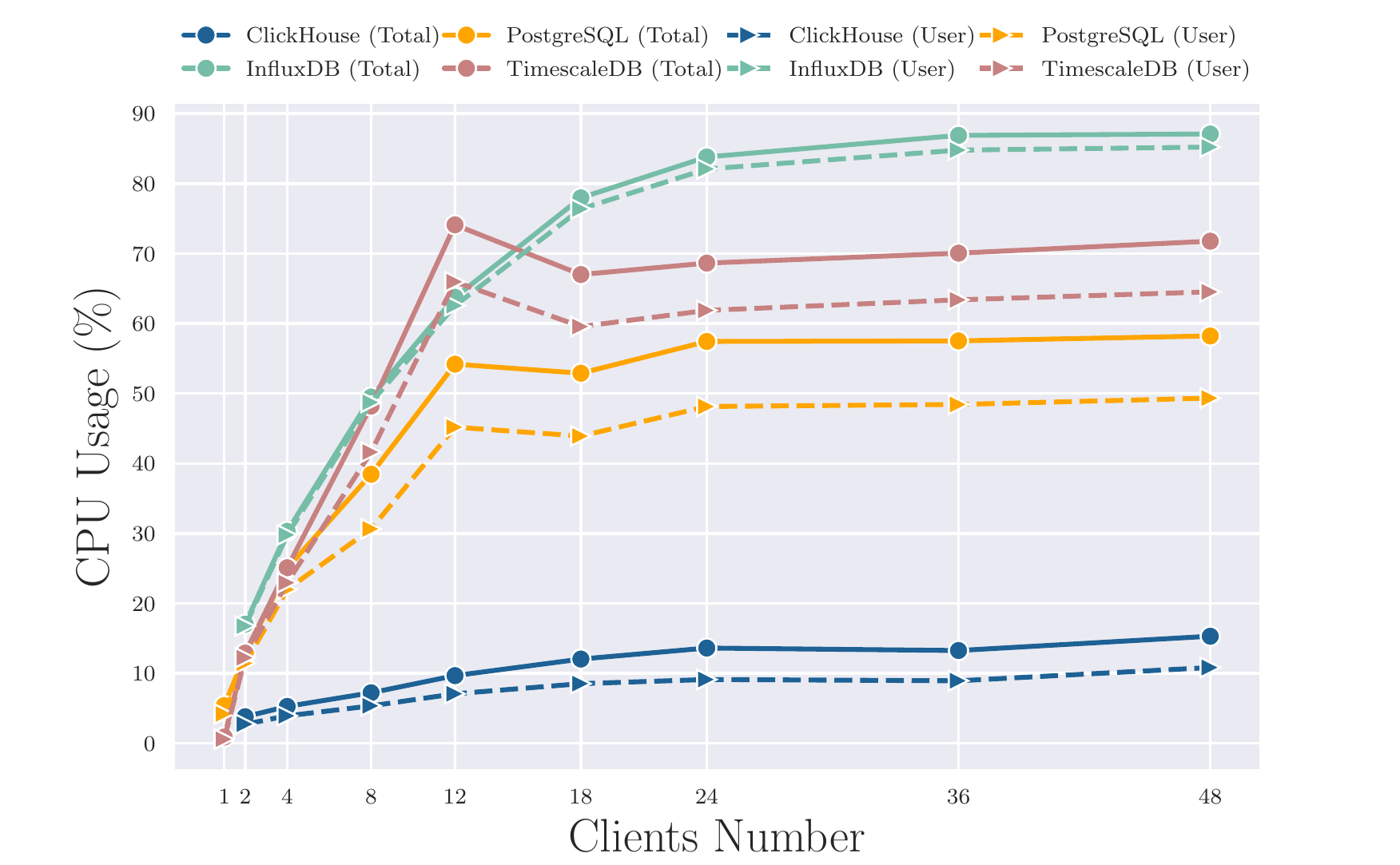}
        \caption{Average CPU Usage}
        \label{fig:clients-cpu}
        \Description{}
    \end{subfigure}
    \hfill
    \begin{subfigure}[b]{0.47\textwidth}
        \centering
        \includegraphics[width=\textwidth]{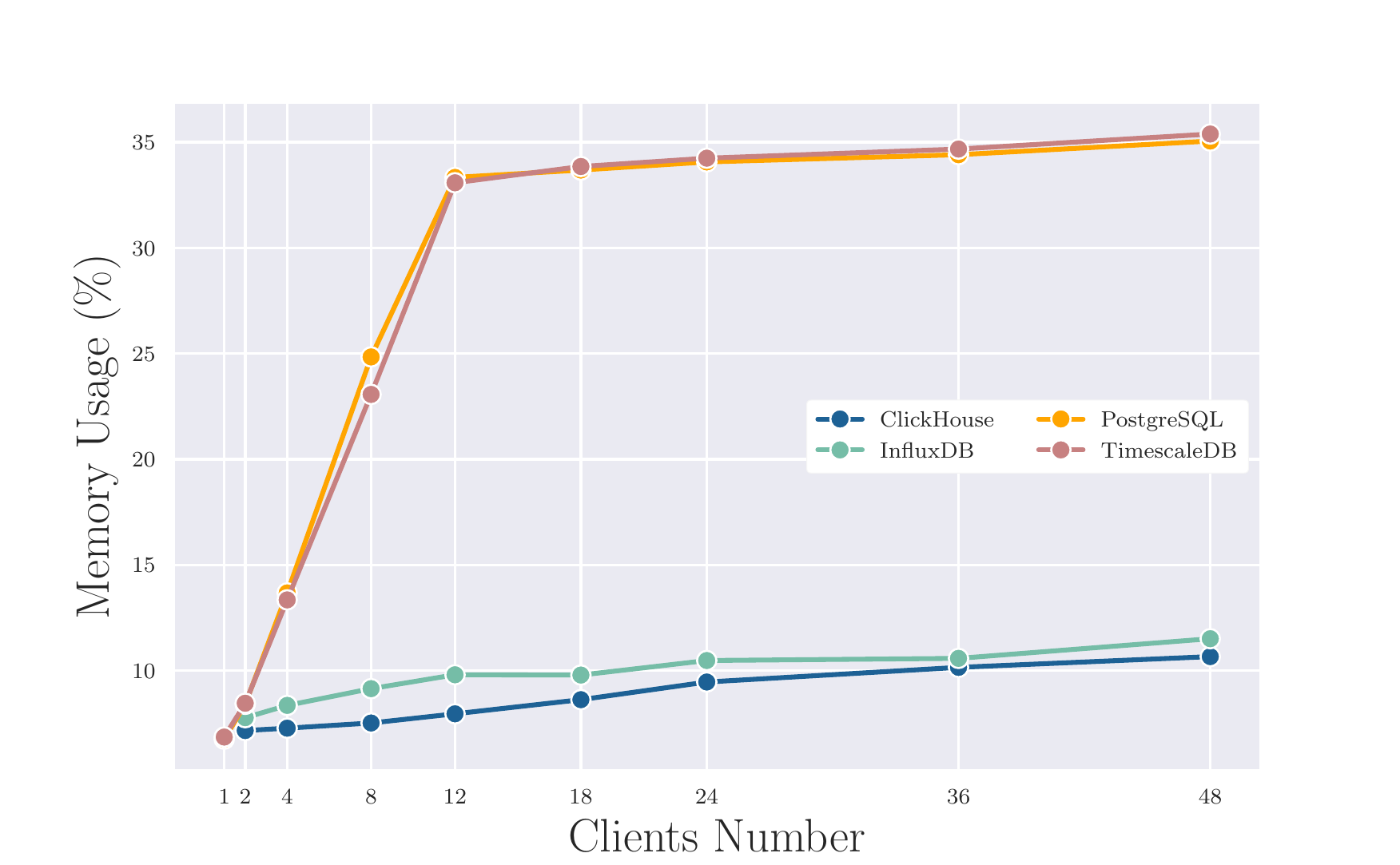}
        \caption{Average Memory Usage}
        \label{fig:clients-mem}
        \Description{}
    \end{subfigure}
    \hfill
    \caption{Usage of System Resources as Function of the Number of concurrent clients}
    \label{fig:clients-metrics}
\end{figure}

\autoref{fig:clients-metrics} shows the usage of system resources as a function of varying the number of concurrent clients. \autoref{fig:clients-cpu} shows average total (solid lines) and user-space (dashed lines) CPU usage per clients number. Although InfluxDB provides a considerably high ingestion rate we notice that its average CPU usage is high even when the number of concurrent clients is below 8. For a higher number of clients, InfluxDB can overload the CPUs. This explains why InfluxDB reached its peak performance at \sepnum{790000} in \autoref{fig:clients}. TimescaleDB and PostgreSQL also show high CPU usage, especially beyond 12 concurrent clients with a wider gap between total CPU usage and userspace usage. The high CPU usage in these two databases is expected as a result of the process forks that are created for each PostgreSQL connection. On the other hand and accompanied by much higher ingestion rates, ClickHouse maintains a considerably low CPU usage even with a large number of concurrent clients.

\autoref{fig:clients-mem} shows the memory usage of the target databases. While InfluxDB and ClickHouse keep a low footprint where they do not exceed more than 15\% as an upper limit even with high numbers of concurrent clients. TimescaleDB and PostgreSQL have the same memory usage footprint and they reach up to 34\% with only 12 concurrent clients.

\paragraph{Scaling Workload} The goal of this workload is to stress and check the performance of the target database server as its size goes larger. We stress each of the target databases with 48 concurrent clients, the maximum number of logical cores the machine \textit{M2} is capable of. Each client continuously inserts batches of \sepnum{20000} records where the ingestion latency is most similar for all databases (as shown in \autoref{fig:batching}) and until we reach around 2.8 billion records in total. \autoref{tab:scaling} shows the ingestion rate and the total time taken to insert around 2.8 billion records in each of the databases. ClickHouse shows the best ingestion performance with the ability to ingest more than 1.2 million records per second then InfluxDB, TimescaleDB, and finally PostgreSQL in order. Compared to PostgreSQL, ClickHouse provides 6x speedup in data ingestion with its OLAP-based design where it writes directly to the storage without passing into leveled write procedures like LSM trees. On the other hand, InfluxDB provides 3.5x speedup in data ingestion using its LSM tree-based storage engine. Being based on PostgreSQL, TimescaleDB inherits some of its limitations and provides only 2.33x speedups in ingestion rate.

\begin{table*}[ht]
    \caption{Total Time, Ingestion Rate, and the Throughput of the Scaling Workload Experiment}
    \label{tab:scaling}
    \begin{tabular}{cccc}
        \toprule
        Target Database & Total Time                                & Ingestion Rate (\SI{}{records/\sec}) & Throughput (\SI{}{\mega\byte/\sec}) \\
        \midrule
        ClickHouse      & \SI{37}{\min} \SI{32}{\sec}               & \sepnum{1278928}                     & \textasciitilde30.69                \\
        InfluxDB        & \SI{1}{\hour} \SI{4}{\min} \SI{43}{\sec}  & \sepnum{741688.5}                    & \textasciitilde17.8                 \\
        TimescaleDB     & \SI{1}{\hour} \SI{37}{\min} \SI{55}{\sec} & \sepnum{490149.8}                    & \textasciitilde11.76                \\
        PostgreSQL      & \SI{3}{\hour} \SI{48}{\min} \SI{10}{\sec} & \sepnum{210361.9}                    & \textasciitilde5.04                 \\
        \bottomrule
    \end{tabular}
\end{table*}

\begin{figure}[ht]
    \centering
    \includegraphics[width=0.46\textwidth]{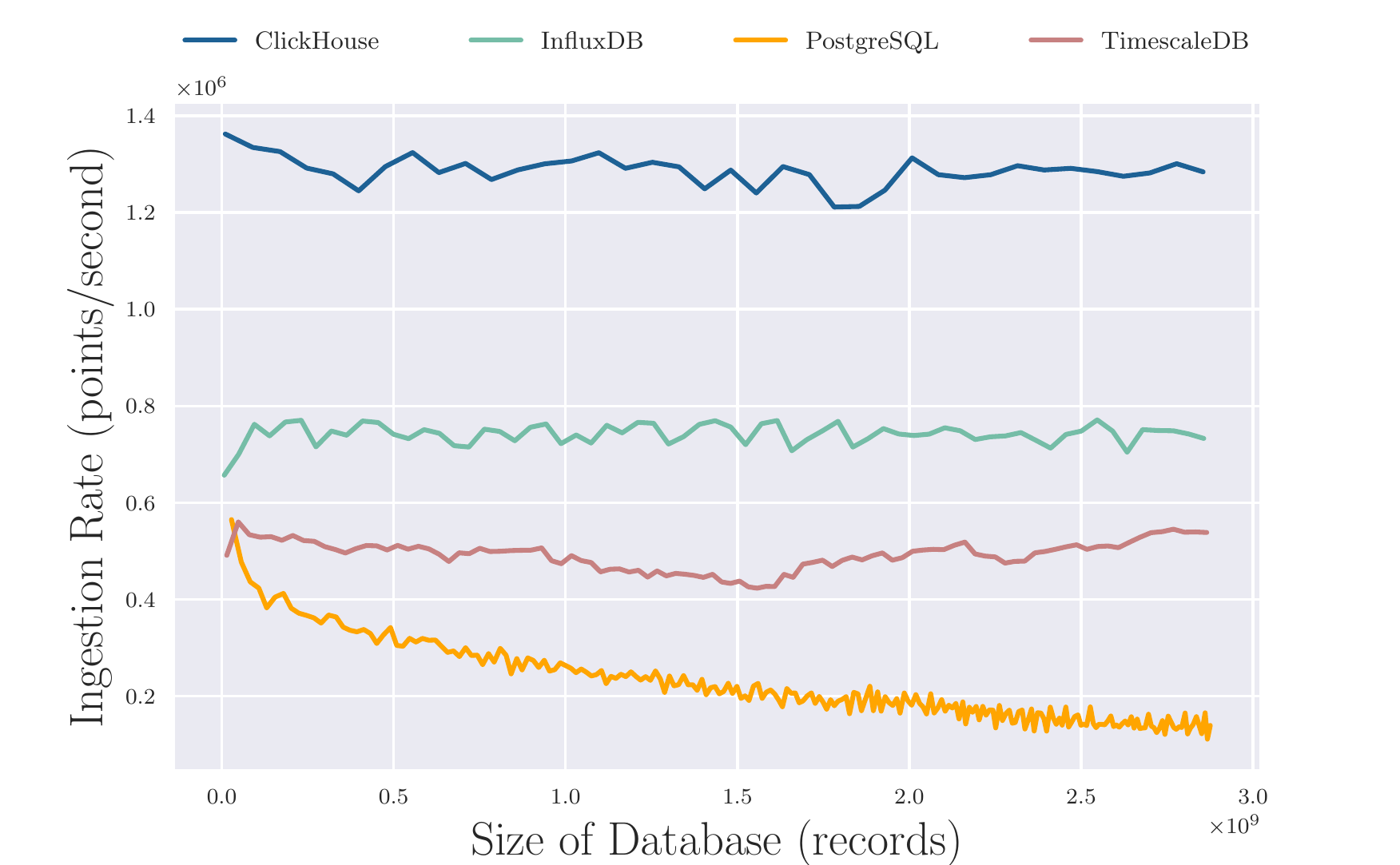}
    \caption{Ingestion Rate (in million records per second) as function of the size of the database.}
    \label{fig:scaling}
    \Description{ClickHouse shows the best ingestion performance with the ability to ingest more than 1.2 million records per second then InfluxDB, TimescaleDB, and finally PostgreSQL. Time-series databases do not only preform much better than PostgreSQL but they also provide stable performance with respect to the table size in the database.}
\end{figure}

\begin{figure}[ht]
    \begin{subfigure}[b]{0.47\textwidth}
        \centering
        \includegraphics[width=\textwidth]{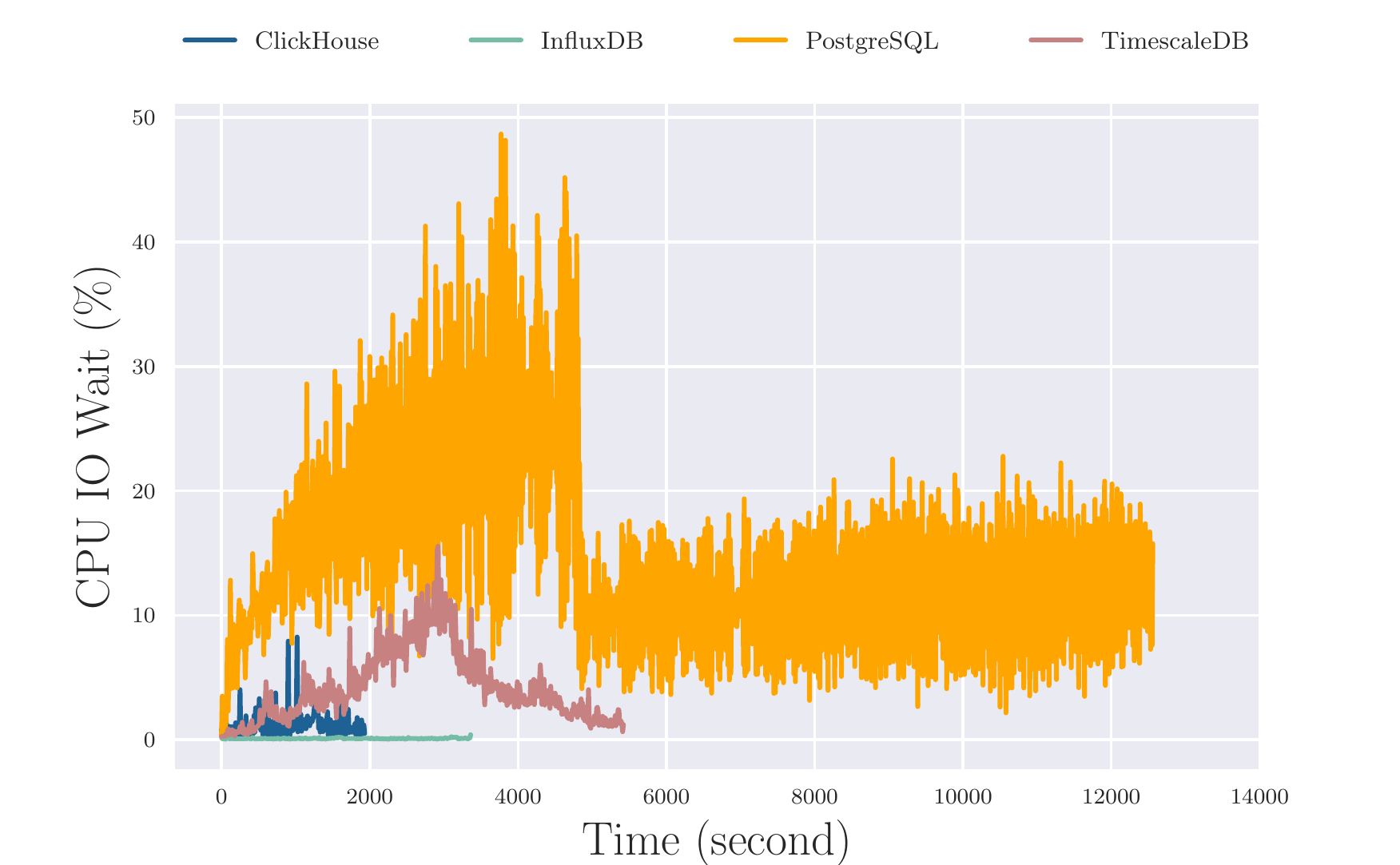}
        \caption{Percentage of CPU Usage of I/O Wait}
        \label{fig:scaling_iowait}
        \Description{The percentage of CPU spent IO Wait is very high for PostgreSQL sever reaching the maximum value around 50\% and averaging around 14.79\%}
    \end{subfigure}
    \hfill
    \begin{subfigure}[b]{0.47\textwidth}
        \centering
        \includegraphics[width=\textwidth]{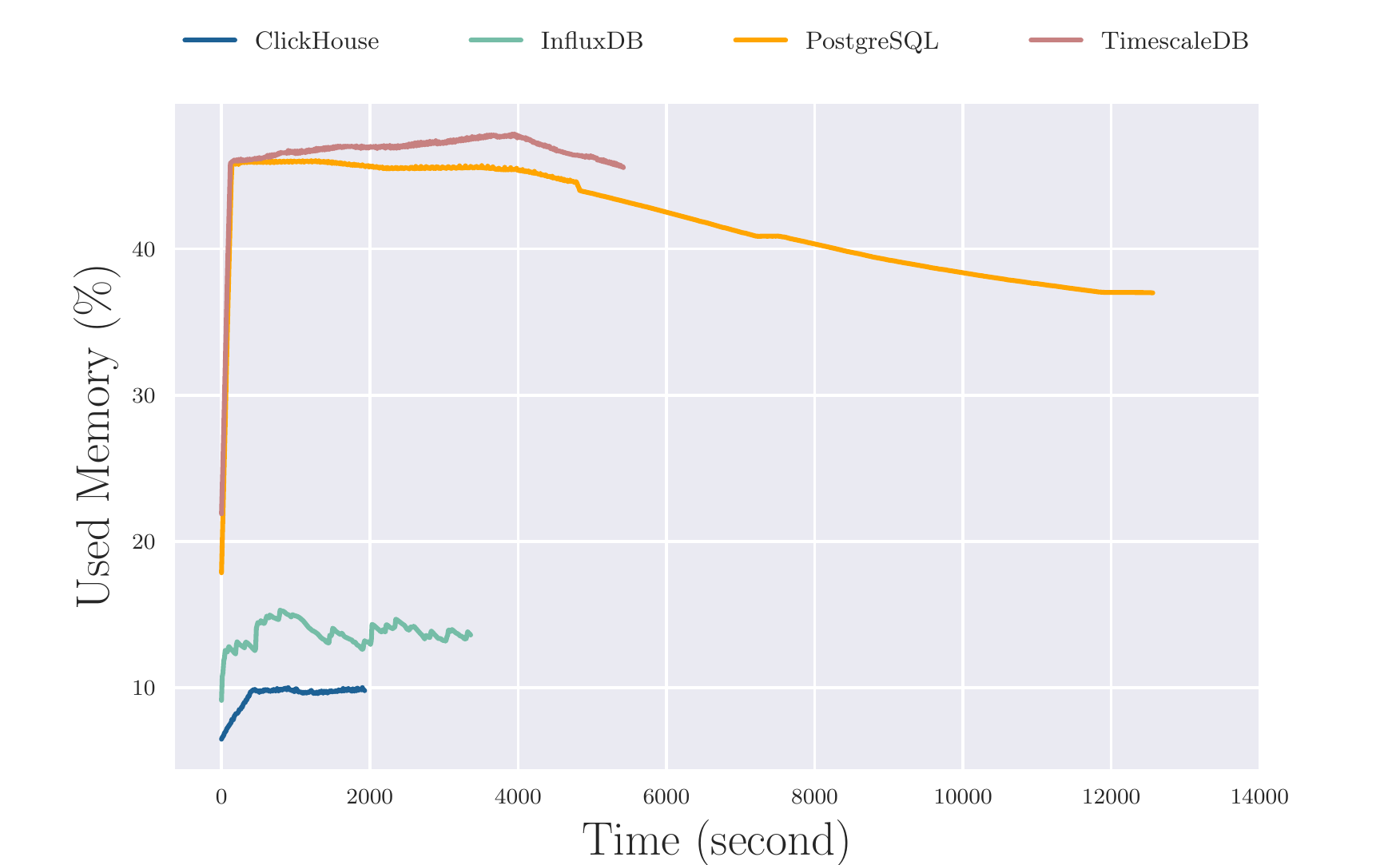}
        \caption{Percentage of Used Physical Memory}
        \label{fig:scaling_usedmem}
        \Description{High memory usage of TimescaleDB and PostgreSQL}
    \end{subfigure}
    \hfill
    \begin{subfigure}[b]{0.47\textwidth}
        \centering
        \includegraphics[width=\textwidth]{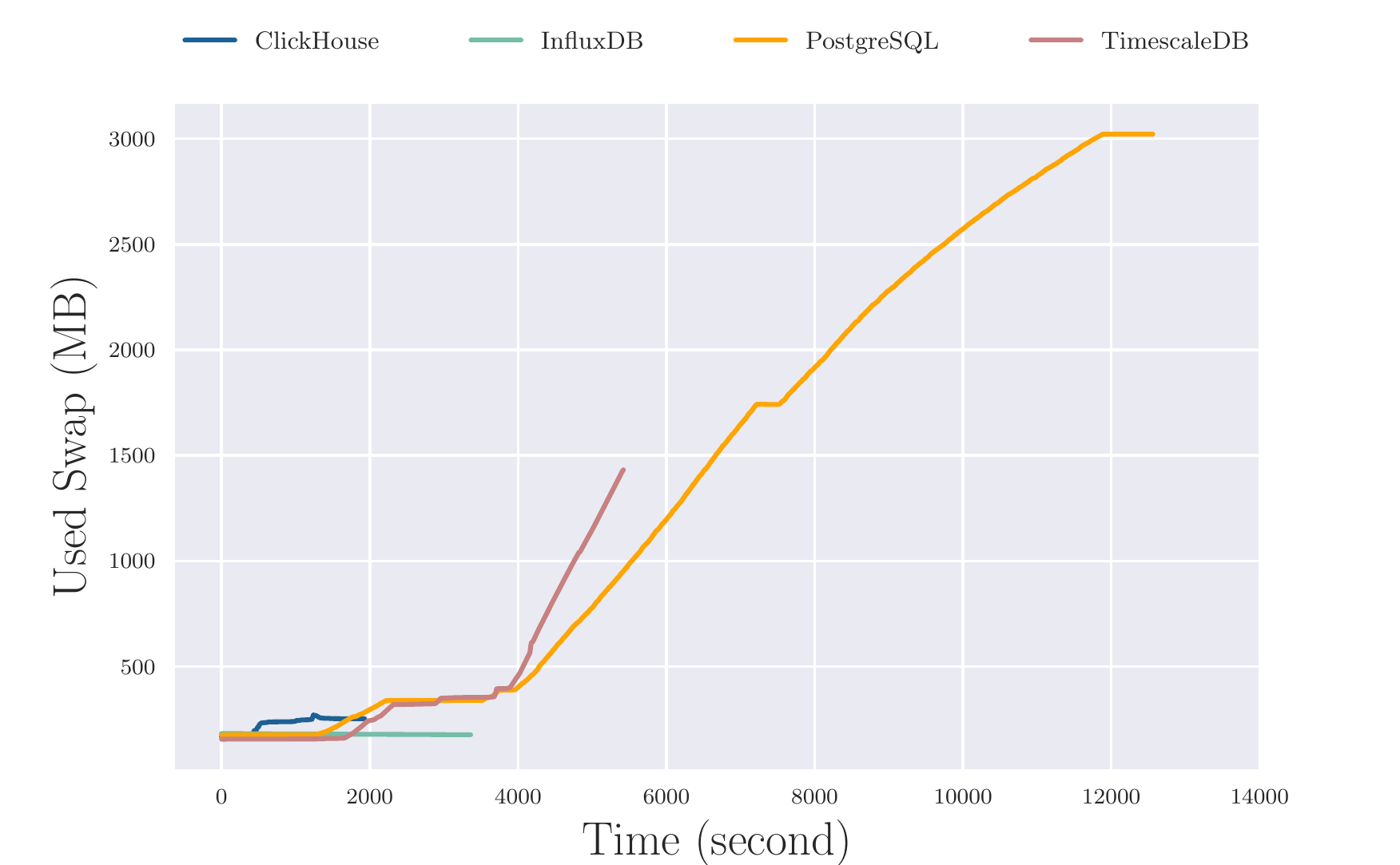}
        \caption{Used Swap Space in Megabytes}
        \label{fig:scaling_swap}
        \Description{Swap is used extensively by PostgreSQL and TimescaleDB}
    \end{subfigure}
    \caption{The Scaling Workload System Metrics for Different Database Servers as Function of Time}
    \label{fig:scaling_metrics}
\end{figure}

\autoref{fig:scaling} shows that time-series databases not only perform much better than PostgreSQL but also provide stable performance with respect to the table size in the database. To understand why the performance of PostgreSQL is dropping we look at its corresponding collected system metrics. \autoref{fig:scaling_metrics} shows the system metrics of the scaling workload for the target database servers as a function of the duration of the experiments. We noticed that the percentage of CPU spent I/O Wait in \autoref{fig:scaling_iowait} is very high for PostgreSQL sever reaching the maximum value around 50\% and averaging around 14.79\%. In addition, \autoref{fig:scaling_usedmem} shows the percentage of used memory of the target database servers. As the data in the database server grows larger, PostgreSQL and TimescaleDB memory usage keep increasing until they reach around 40\% and the operating system starts swapping database pages to the storage disks as shown in \autoref{fig:scaling_swap}. On the other hand, InfluxDB and ClickHouse use up to 20\% of the physical memory with a negligible swap usage. PostgreSQL's ingestion rate performance degradation is caused by swapping indexes in and out from the physical memory as the time-series data in the database grows larger. TimescaleDB solves these shortcomings of PostgreSQL by optimizing the usage of the physical memory through chunking the big table to partitions whose indexes are independent and can fit into the physical memory, thus it does not rely on the swap as PostgreSQL does.

\subsection{Queries Latency}
We discuss the performance of \scits{} queries. We fill the database with 2.8 billion records that correspond to a duration of 15 days and for \sepnum{100000} sensors. For each query, we execute 20 runs. For each query run, we clear the database tables and the operating system caches, and restart the database server to make sure the query results are directly returned from disk and choose distinct parameters.

\paragraph{Q1. Raw Data Fetching}
It queries the database to read the time-series data of a \SI{10}{\min} duration for 10 distinct sensors. Each \SI{10}{\min} interval is randomly selected from 15-day dataset using uniform distribution. The duration corresponds to around \sepnum{5000} data points. \autoref{tab:q1} shows the query latency statistics in milliseconds for Q1. The query latency is lowest on ClickHouse where it records \SI{272}{\ms} as a maximum value and \SI{177.7}{\ms} as an average value. PostgreSQL with its B-Tree indexing is second in performance with \SI{457}{\ms} as a maximum value and \SI{361.7}{\ms} on average. InfluxDB is third with \SI{1172}{\ms} as a maximum value and \SI{1352}{\ms} in average and greater deviation than that of ClickHouse and PostgreSQL. The disadvantages of chunking a table are realized when the TimescaleDB is forth with \SI{1352}{\ms} as a maximum value and \SI{1284.55}{\ms} in average and with the greatest deviation.

\begin{table}[ht]
    \caption{Query Latency Statistics (in ms) for Q1}
    \label{tab:q1}
    \begin{tabular}{lccccc}
        \toprule
        Database    & Min. & Mean   & 95\%    & Max. & Std. Dev. \\
        \midrule
        ClickHouse  & 131  & 177.7  & 241.6   & 272  & 32.64     \\
        InfluxDB    & 567  & 737.5  & 1058.95 & 1172 & 161.36    \\
        TimescaleDB & 608  & 910.75 & 1284.55 & 1352 & 217.57    \\
        PostgreSQL  & 283  & 361.7  & 426.6   & 457  & 51.64     \\
        \bottomrule
    \end{tabular}
\end{table}

\paragraph{Q2. Out of Range}
We query the database for the day hours where the data of exactly one sensor is considered out of range according to user-defined boundaries in a duration of \SI{180}{\min} of time-series of data. The \SI{180}{\min} duration is randomly selected from the 15-day dataset using uniform distribution. \autoref{tab:q2} shows the query latency statistics in milliseconds for Q2. ClickHouse again achieves first place with a maximum value of \SI{263}{\ms} and an average value of \SI{188.35}{\ms}. TimescaleDB comes in second with a maximum value of \SI{602}{\ms} and \SI{440.3}{\ms} average value. InfluxDB achieves very similar performance to TimescaleDB but with a maximum value of \SI{627}{\ms} and a \SI{442.35}{\ms} average value. With complex queries like Q2, PostgreSQL starts to show some performance limitations where the maximum value records \SI{1950}{\ms} and the average value records \SI{1707.15}{\ms}.

\begin{table}[ht]
    \caption{Query Latency Statistics (in ms) for Q2}
    \label{tab:q2}
    \begin{tabular}{lccccc}
        \toprule
        Database    & Min. & Mean    & 95\%   & Max. & Std. Dev. \\
        \midrule
        ClickHouse  & 142  & 188.35  & 219.3  & 263  & 26.04     \\
        InfluxDB    & 387  & 442.35  & 512.05 & 627  & 54.96     \\
        TimescaleDB & 314  & 440.3   & 544.05 & 602  & 81.61     \\
        PostgreSQL  & 1539 & 1707.15 & 1779   & 1950 & 90.71     \\
        \bottomrule
    \end{tabular}
\end{table}

\paragraph{Q3. Data Aggregation}
We query the database to calculate the standard deviation of the values of 10 sensors over a \SI{60}{\min} time interval. The \SI{60}{\min} duration is randomly selected from 15-day dataset using uniform distribution. \autoref{tab:q3} shows the query latency statistics in milliseconds for Q3. ClickHouse ranks first in Q3 performance with a maximum value of \SI{244}{\ms} and \SI{203.55}{\ms} average value. InfluxDB performs better than TimescaleDB with a maximum value of \SI{594}{\ms} and \SI{427.85}{\ms} average value while TimescaleDB records \SI{791}{\ms} as a maximum value and \SI{571.95}{\ms} average value, but TimescaleDB records a high standard deviation and lower minimum value than InfluxDB. PostgreSQL records the least performance for Q4 with a maximum value of \SI{763}{\ms} and an average value of \SI{657.4}{\ms}.

\begin{table}[ht]
    \caption{Query Latency Statistics (in ms) for Q3}
    \label{tab:q3}
    \begin{tabular}{lccccc}
        \toprule
        Database    & Min. & Mean   & 95\%   & Max. & Std. Dev. \\
        \midrule
        ClickHouse  & 167  & 203.55 & 238.3  & 244  & 22.33     \\
        InfluxDB    & 280  & 427.85 & 555.05 & 594  & 69.04     \\
        TimescaleDB & 268  & 571.95 & 691.25 & 791  & 106.54    \\
        PostgreSQL  & 600  & 657.4  & 737.35 & 763  & 47.12     \\
        \bottomrule
    \end{tabular}
\end{table}

\paragraph{Q4. Data Downsampling}
We query the database to summarize the data of 10 sensors over \SI{24}{\hour} every hour. The \SI{24}{\hour} duration is randomly selected from 15-day dataset using uniform distribution. \autoref{tab:q4} shows the query latency statistics in milliseconds for Q4. Even with a complex query like Q4, ClickHouse is still ranking first with a maximum value of \SI{300}{\ms} and \SI{293.35}{\ms} average value. InfluxDB and TimescaleDB give a reasonable performance with a maximum value of \SI{873}{\ms} and \SI{647.9}{\ms} average value for InfluxDB while \SI{1024}{\ms} as a maximum value and \SI{754.6}{\ms} average value for TimescaleDB. With a complex query like Q4, PostgreSQL records a bad performance that is ranging between \SI{9858}{\ms} and \SI{14157}{\ms} and averaging at \SI{13445.95}{\ms}.

\begin{table}[ht]
    \caption{Query Latency Statistics (in ms) for Q4}
    \label{tab:q4}
    \begin{tabular}{lccccc}
        \toprule
        Database    & Min. & Mean     & 95\%    & Max.  & Std. Dev. \\
        \midrule
        ClickHouse  & 175  & 237.45   & 293.35  & 300   & 33.42     \\
        InfluxDB    & 464  & 647.9    & 816     & 873   & 87.15     \\
        TimescaleDB & 548  & 754.6    & 965.1   & 1024  & 114.35    \\
        PostgreSQL  & 9858 & 13445.95 & 13974.6 & 14157 & 894.96    \\
        \bottomrule
    \end{tabular}
\end{table}

\paragraph{Q5. Operations on Two Down-sampled Sensors}
We query the database to calculate the difference between the summarized data of two sensors over \SI{24}{\hour}. The data is summarized every one hour that is randomly selected from 15-day dataset using uniform distribution. \autoref{tab:q5} shows the query latency statistics in milliseconds for Q5. ClickHouse records the best performance with a maximum value of \SI{419}{\ms} and an average of \SI{301.7}{\ms}. TimescaleDB outperforms InfluxDB in this query with a maximum value of \SI{701}{\ms} and \SI{448.6}{\ms} on average for TimescaleDB while a maximum value of \SI{810}{\ms} and \SI{522.4}{\ms} on average for InfluxDB. PostgreSQL does not perform well for complex queries, it records very high latencies that are \SI{20806.15}{\ms} on average.

\begin{table}[ht]
    \caption{Query Latency Statistics (in ms) for Q5}
    \label{tab:q5}
    \begin{tabular}{lccccc}
        \toprule
        Database    & Min.  & Mean    & 95\%    & Max.  & Std. Dev. \\
        \midrule
        ClickHouse  & 167   & 301.7   & 397.15  & 419   & 68.43     \\
        InfluxDB    & 430   & 522.4   & 779.6   & 810   & 109.49    \\
        TimescaleDB & 209   & 448.6   & 666.8   & 701   & 138.29    \\
        PostgreSQL  & 20344 & 20806.1 & 21134.8 & 21151 & 236.76    \\
        \bottomrule
    \end{tabular}
\end{table}

ClickHouse provides an outstanding stable query performance because of its unique data storage. In addition to its foundational columnar format, ClickHouse partitions data in multiple files and uses a sparse indexing algorithm where indexes are stored for every N-th row of the table instead of indexing every row which supports querying data in ranges as is the case of time-series data. Even for complex queries like Q4 and Q5, ClickHouse provides very good performance without being impacted because of the performance benefits of cross-breeding vectorized query execution and compiled query execution \cite{query_execution}. InfluxDB and TimescaleDB offer close performance while their backends are different but with conceptual similarities. InfluxDB uses the columnar format and a multi-level indexing mechanism where a query starts by determining in which partition/file the time range is, once the partition and its corresponding files are determined, InfluxDB does a binary search to find the requested data. On the other hand, TimescaleDB is row-based for recent data, but if compression is enabled, it uses a row-column hybrid model where the columns of multiple are stored in separate arrays. TimescaleDB queries start by determining which chunks have the requested data then it uses B-Tree indexes of this chunk to determine which rows have the data. The clear disadvantages of a complete row-based model and the absence of data partitioning are present with PostgreSQL.

\section{Related Work}
\label{sec:relatedwork}

Understanding the performance of databases has been a topic of interest for so long. Performance evaluation of databases helps in capacity planning and in choosing the most suitable database for a specific use case like time-series data workloads, big data workloads, or transaction-based workloads. The most notable benchmarks are the benchmarks from the TPC council for OLTP databases e.g. TPC-C, TPC-DS, and TPC-H. The scientific community also introduced other benchmarks like \cite{oltpbenchmark} for OLTP databases or YCSB \cite{ycsb} for big data databases.

TPCx-IoT is the IoT benchmark from the TPC council. Its workloads simulate data from energy power plants in the form of data ingestion and concurrent queries. TPCx-IoT supports very basic queries which makes it not suitable for many practical uses. TSBS \cite{tsbs} is a benchmark from the developers of the TimescaleDB company. TSBS simulates a load of IoT devices in addition to DevOps, but TSBS lacks concurrency and the ability to read the usage of system resources. Rui Lui et al. propose the IoTDB-Benchmark \cite{iotdbbenchmark} for IoT scenarios. IoTDB-Benchmark supports concurrent, aggregation, and down-sampling queries. YCSB-TS \cite{ycbsts} adopts the structure and the workloads of YCSB and adds basic time functions and thus inherits unoptimized workloads to benchmark time-series databases. \textit{ts-benchmark} \cite{tsbenchmark} is a time-series benchmark developed by Yuanzhe Hao et al. It uses a generative adversarial network (GAN) model to generate synthetic time-series data to ingest data and supports diverse workloads for data loading, injection, and loading in addition to monitoring usage of system resources. \textit{ts-benchmark}, however, does not take into consideration aggregation and down-sampling queries which are very important for data visualization and analysis.

\section{Conclusion}
\label{sec:conclusion}
Although SciTS is inspired by scientific experiments and industrial IoT, it is a highly flexible benchmark that can cover most of the ingestion workloads through parameterization of sensors cardinality, concurrency, and size of the inserted batch.
We also introduce the "Scaling Workload", a novel workload to study the performance of the database as its size grows larger. \scits{} embeds 5 data mining queries inspired by the requirements of data management and data analysis systems of currently operating large-scale scientific facilities. These tests characterize performance of including range, aggregation, down-sampling, and statistical queries.

We evaluate the performance of the 4 databases with 4 distinct designs using \scits{}: ClickHouse as a completely column-oriented OLAP DBMS, InfluxDB as an LSM-Tree database, and TimescaleDB as an ACID database with adaptations to time-series data, and PostgreSQL to represent traditional DBMS. We demonstrate the advantages of time-series databases as compared to traditional relational databases using PostgreSQL as an example. Our evaluation shows that the selected TSDBs outperform PostgreSQL up to 6-times and, unlike PostgreSQL preserves, stable ingestion rate over time. Even higher performance advantage is achieved in the queries aimed to model data mining and analysis workloads typical for large-scale scientific experiments. To help readers with selection and configuration of appropriate time-series database for their use-cases, we further discuss the impact of the database design on its performance.Our results on the targeted databases can be summarized as follows:

\paragraph{ClickHouse} With its simple but efficient storage engine, ClickHouse supports very high ingestion rates up to 1.3 million records per second in addition to very efficient concurrency handling. Even for high ingestion rates and up to 48 concurrent clients as far as we tested, ClickHouse keeps low CPU usage and a very low memory footprint. ClickHouse significantly outperforms other evaluated databases in the speed of data queries and shows reasonably low deviation in query latency. Queries of ClickHouse are powered by its partitioned data management, a sparse indexing algorithm, and a very efficient mix of compiled and vectorized query execution.

\paragraph{InfluxDB} The LSM-Tree design of InfluxDB produces a relatively high ingestion rate with up to around \sepnum{790000} records per second. However, this is how far InfluxDB can go with our current hardware setup. Although InfluxDB is hungry for CPU resources, it is very light on memory due to its efficient and diverse data compression algorithms. Performance of data queries is second to ClickHouse and the database engine also shows low variability in latency.

\paragraph{TimescaleDB} Compared to PostgreSQL, TimescaleDB is a huge improvement. Its design tries to prove that the ACID principles can still hold the high ingestion rate of time-series data. With our setup, we can achieve
ingestion rate of \sepnum{490000} records per second. Compared to PostgreSQL, it optimizes the usage of system resources with efficient CPU and virtual memory usage. TimescaleDB also significantly improves latency of all evaluated data
queries over standard PostgreSQL performance.

\paragraph{PostgreSQL} The traditional ACID design of PostgreSQL fails to maintain data over the long run with degrading write performance caused by maintaining very large indexes in the virtual memory. PostgreSQL's ingestion rate is very small compared to TSDB databases and usage of system resources is not efficient with high CPU usage.

\scits{} shows that the unique designs of time-series databases bring outstanding performance benefits in addition to the easy management and manipulation of big time-series data compared to traditional databases. We see the benefits of relaxing the consistency constraints for performance. The columnar format of databases and in addition to data partitioning into multiple parts boosts TSDB ingestion rates and leads to improved performance of data queries. On the side of the system performance, time-series databases are lighter on the system resources with very efficient memory management. \scits{} showed the importance of TSDB in managing and storing time-series data on a single node. We are looking to extend \scits{} to support clustered multi-node database servers setups.

\clearpage
\balance
\bibliographystyle{ACM-Reference-Format}
\bibliography{tsdb}

\end{document}